\documentclass[prl,aps,twocolumn,floatfix,showpacs]{revtex4}
\usepackage{graphics}
\begin{document}
\title{Ultracold heteronuclear molecules and ferroelectric superfluids}
\author{M. Iskin and C. A. R. S{\'a} de Melo}
\affiliation{School of Physics, Georgia Institute of Technology, Atlanta, Georgia 30332, USA}
\date{\today}

\begin{abstract}

We analyze the possibility of a ferroelectric transition in heteronuclear molecules consisting
of Bose-Bose, Bose-Fermi or Fermi-Fermi atom pairs. This transition is 
characterized by the appearance of a spontaneous electric polarization 
below a critical temperature. 
We discuss the existence of a ferroelectric Fermi liquid phase for Fermi molecules
and the existence of a ferroelectric superfluid phase for Bose molecules
characterized by the coexistence of ferroelectric and superfluid orders.
Lastly, we propose an experiment to detect ferroelectric correlations through 
the observation of coherent dipole radiation pulses during time of flight.

\end{abstract}
\maketitle

Arguably one of the next frontiers in ultracold atomic and molecular physics
is the study of two-species atomic 
mixtures~\cite{truscott, schrenk, hadzibabic, roati, modugno, silber},
and ultracold heteronuclear molecules composed of two-species alkali atoms such as 
KRb~\cite{mancini, wang, bongs}, RbCs~\cite{kerman} and NaCs~\cite{haimberger}.
This frontier may be advanced through the use of Feshbach resonances which have 
already been observed in mixtures of two-species alkali atoms~\cite{stan, inouye, ferlaino}, 
and may also become a crucial tool for tuning physical properties of heteronuclear
systems.

Ultracold heteronuclear molecules made of Bose-Bose, Bose-Fermi 
or Fermi-Fermi atom pairs offer many new opportunities compared to 
standard (Bose or Fermi) atomic systems because of their 
additional degrees of freedom.
For instance, when heteronuclear diatomic molecules are formed from neutral atoms, 
electric charge is transferred from one atom to the other
leading to an electric dipole moment $|\mathbf{p}| = Q(d) d$,
where $d$ is the separation and $Q(d)$ is the effective charge 
transfer between constituent atoms~\cite{aymar}. These electric dipoles 
have equal magnitudes but random orientations at high temperatures 
leading to a vanishing average electric polarization. 
However, at low temperatures, the dipoles may all point to a particular direction 
producing a finite average electric polarization density $\langle {\bf P} \rangle$, 
characteristic of a ferroelectric state. In addtion, when ultracold heteronuclear 
molecules form a Bose-Einstein condensate (BEC) a ferroelectric 
superfluid state proposed in this manuscript may be accessible experimentally.
\begin{figure} [htb]
\centerline{\scalebox{0.4}{\includegraphics{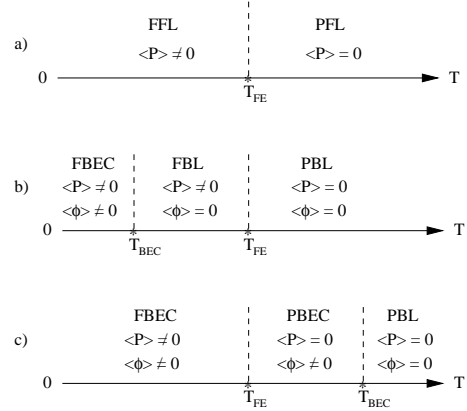} } }
\caption{\label{fig:one}
Schematic phase diagram for polar a) Fermi, and b), c) Bose molecules,
where
$\langle \mathbf{P} \rangle$ and $\langle \phi \rangle$ 
are the corresponding order parameters for ferroelectric
order at $T_{FE}$, and for BEC at $T_{BEC}$, respectively.
}
\end{figure}

Our main results are as follows.
When heteronuclear molecules are composite fermions (Bose-Fermi pairs), a phase transition
occurs separating a paraelectric Fermi liquid (PFL) from a ferroelectric Fermi liquid (FFL),
as shown in Fig.~\ref{fig:one}(a).
These phases do not exist in standard condensed matter systems, since all 
experimentally known ferroelectrics are very good insulators~\cite{shirane}, and are not Fermi liquids.
Furthermore, when heteronuclear molecules are composite bosons (Bose-Bose or Fermi-Fermi pairs),
a ferroelectric transition with critical temperature $T_{FE}$ may occur either 
above or below the BEC temperature $T_{BEC}$. 
When the molecular dipole moment and/or density 
are sufficiently large then $T_{FE} > T_{BEC}$ leading
to a paraelectric Bose liquid (PBL) for $T > T_{FE}$, a ferroelectric Bose liquid (FBL)
for $T_{FE} > T >  T_{BEC}$, and to a ferroelectric BEC (FBEC) for $T < T_{BEC}$, 
as shown in Fig.~\ref{fig:one}(b).
When the molecular dipole moment and/or density are sufficiently low then
$T_{FE} < T_{BEC}$ leading to a paraelectric Bose liquid (PBL) for $T > T_{BEC}$, 
a paraelectric BEC for $T_{BEC} > T >  T_{FE}$, and to a ferroelectric BEC (FBEC) for $T < T_{FE}$, 
as shown in Fig.~\ref{fig:one}(c). The FBEC phase corresponds to a ferroelectric superfluid.

{\it Hamiltonian:} To describe all these phases of ultracold heteronuclear molecules, we use
the Hamiltonian density 
\begin{eqnarray}
\label{eqn:first-hamiltonian}
H(\mathbf{r}) &=& \sum_\sigma \psi_\sigma^*(\mathbf{r}) \left[-\frac{\nabla^2} {2M} 
- \mu + V(\mathbf{r}) \right] \psi_\sigma( \mathbf {r}) \nonumber \\
&+& \frac{1}{2} \sum_{\sigma,\sigma'} \int d\mathbf{r'} n_{\sigma'} (\mathbf{r}) 
U_d(\mathbf{r}, \mathbf{r'}) n_{\sigma}(\mathbf{r}) \nonumber \\
&-& \frac{1}{2} \sum_{\alpha,\beta} \int d\mathbf {r'} \mathbf{P}_\alpha(\mathbf {r'}) 
J_{\alpha, \beta}(\mathbf{r}, \mathbf{r'}) \mathbf{P}_\beta(\mathbf{r}),
\end{eqnarray}
describing a weakly interacting gas of dilute polar molecules,  
where $\psi_\sigma^*(\mathbf {r})$ creates a Bose (Fermi) molecule at position $\mathbf {r}$ 
with pseudo-spin-$\sigma$,
and $n_\sigma(\mathbf{r}) = \psi_\sigma^*(\mathbf {r}) \psi_\sigma(\mathbf {r})$ is the density operator.
The first term in Eq.~(\ref{eqn:first-hamiltonian}) 
is the kinetic energy, where $\mu$ is the chemical potential, 
and $V (\mathbf{r})$ is the trapping potential.
The second term is the density-density interaction between Bose (Fermi) 
molecules, and for a contact interaction is given by 
$
U_d(\mathbf{r}, \mathbf{r^\prime}) = U \delta(\mathbf{\Delta r}),
$
where 
$ U = 4\pi a / M $
is the amplitude of the interaction and $\delta(\mathbf{\Delta r})$ is the delta function
with $\mathbf{\Delta r} = \mathbf{r} - \mathbf{r^\prime}$.
Here, $a$ is the scattering length of the corresponding Bose (Fermi) molecules,
and $M$ is the molecular mass.
The third term is the electric dipole-dipole interaction 
between molecular dipoles at positions $\mathbf{r}$ and $\mathbf{r^\prime}$, 
where
$
\mathbf{P}(\mathbf{r}) = \mathbf{p} (\mathbf{r}) \sum_\sigma n_\sigma(\mathbf{r})
$
is the polarization operator, and $\mathbf{p} (\mathbf{r})$ is the 
molecular dipole moment at position $\mathbf{r}$.
The electric dipole-dipole interaction is given by
$
J(\mathbf{r}, \mathbf{r^\prime}) = \sum_{\alpha, \beta} \mathbf{p}_\alpha (\mathbf{r}) 
J_{\alpha, \beta} (\mathbf{r}, \mathbf{r^\prime}) \mathbf{p}_\beta (\mathbf{r^\prime}),
$
where
\begin{equation}
J_{\alpha, \beta} (\mathbf{r}, \mathbf{r^\prime}) = \frac{1}{4\pi\varepsilon_0} 
\frac{3 [\mathbf{\Delta r}_\alpha \mathbf{\Delta r}_\beta] / |\mathbf{ \Delta r}|^2 - \delta_{\alpha, \beta}} {|\mathbf{\Delta r}|^3}.
\end{equation}
Here, $\delta_{\alpha, \beta}$ is the Kronecker-delta and 
$\lbrace \alpha,\beta \rbrace = \lbrace 1,2,3 \rbrace$ label the vector components.

{\it Electric polarization:} Within the Hartree-Fock approximation, the Hamiltonian density
reduces to 
\begin{eqnarray}
\label{eqn:hamiltonian}
H_0 (\mathbf{r}) &=& - \frac{\nabla^2} {2M} - \widetilde{\mu} + V(\mathbf{r}) + U n_0(\mathbf{r}) 
\nonumber \\
&-& \sum_{\alpha,\beta} \int d\mathbf{r'} \mathbf{p}_\alpha (\mathbf{r})
J_{\alpha, \beta} (\mathbf{r}, \mathbf{r'}) \mathbf{P}_{0, \beta} (\mathbf{r'}),
\end{eqnarray}
where the electric polarization density
$\mathbf{P}_{0}(\mathbf{r}) = \mathbf{p} ({\bf r}) n_0 (\mathbf{r})$,
with 
$
n_0 (\mathbf{r}) = \sum_{\sigma} n_{0,\sigma} (\mathbf{r})
$
being the local density of Bose (Fermi) molecules at $\mathbf{r}$.
The spatially averaged polarization 
$
\langle \mathbf{P} \rangle  = \int d\mathbf{r} \mathbf{P}_{0}(\mathbf{r})/V_c
$
can be rewritten as
\begin{equation}
\langle \mathbf{P} \rangle  =  \frac {1} {V_c} \int d\mathbf{r} \mathbf{p} 
(\mathbf{r}) \sum_{i,\sigma} |\phi_i(\mathbf{r})|^2 f_{\eta} (\epsilon_i),
\label{eqn:polarization}
\end{equation}
where $V_c$ is the sample volume, $\phi_i(\mathbf{r})$ and $\epsilon_i$ are eigenfunctions and eigenvalues of 
the Hamiltonian $H_0(\mathbf{r})$, and $f_{\eta}(\epsilon_i) = 1/[e^{\beta \epsilon_i} - \eta]$ 
is the Bose (Fermi) distribution for Bose (Fermi) molecules when $\eta = 1$ $(-1)$.
The solution of Eq.~(\ref{eqn:polarization}) is non-trivial, however
analytical insight can be gained for homogenous systems where $V(\mathbf{r}) = 0$
and $\mathbf{P}_{0}(\mathbf{r}) = \mathbf{P}_{0}$ is independent of $\mathbf{r}$.
We discuss the homogeneous case first and then analyze the case of a harmonic trap.

{\it Ferroelectric critical temperature:} In the ferroelectric state,
all molecular dipoles are pointing along the same direction $\widehat{\mathbf{m}}$ such that
$
\mathbf{p} (\mathbf{r}) = \mathbf{p} = |\mathbf{p}| \widehat{\mathbf{m}}
$
and $\mathbf{P}_{0} = |\mathbf{P}_0| \widehat{\mathbf{m}}$.
The critical temperature $T_{FE}$ for the ferroelectric transition
is found from the slope of Eq.~(\ref{eqn:polarization}) 
with respect to $\mathbf{P}_{0}$ evaluated at $\mathbf{P}_{0} = 0$, leading to
\begin{equation}
1 + \widetilde{J} (\mathbf{q} \to \mathbf{0}) \left( \frac{\partial N}{\partial \mu} \right)_T = 0,
\label{eqn:tc-first}
\end{equation}
from which $T_{FE}$ can be calculated.
Here, $\widetilde{J} (\mathbf{q})$ with $\mathbf{q} = \mathbf{k} - \mathbf{k'}$ 
is the Fourier transform of 
$J (\mathbf{\Delta r})$, and $N$ is the number of Bose or Fermi molecules. 
Introducing the Kac parameter $\gamma$ as 
a cutoff for short distances ($|\mathbf{\Delta r}| < 1 / \gamma$) leads to
$
\widetilde{J}(\mathbf{q}) = \sum_{\alpha, \beta} \mathbf{p}_{\alpha} 
\widetilde{J}_{\alpha, \beta} (\mathbf{k}, \mathbf{k'}) \mathbf{p}_{\beta},
$
where
\begin{equation}
\widetilde{J}_{\alpha,\beta} (\mathbf{k}, \mathbf{k'}) = 
\frac{S(\widetilde{q})}{\varepsilon_0} \left( \frac{3\mathbf{q}_\alpha \mathbf{q}_\beta} 
{|\mathbf{q}|^2} - \delta_{\alpha,\beta} \right).
\end{equation}
Here, 
$
S(\widetilde{q}) = \sin \widetilde{q} / \widetilde{q}^3 - \cos \widetilde{q} / \widetilde{q}^2
$
where
$\widetilde{q} = |\mathbf{q}| / \gamma$.
Thus, we obtain an implicit relation for the $T_{FE}$
\begin{equation}
1 - \frac{2|\mathbf{p}|^2} {3 \varepsilon_0} n^2 \kappa(T_{FE}) = 0,
\label{eqn:tc-general}
\end{equation}
in terms of the molecular density $n = N/V_c$ and the
isothermal compressibility 
$
\kappa(T) = (1/n^2) (\partial n / \partial \mu)_T.
$ 

The ferroelectric instability is accompanied by a divergence of the dielectric function 
in the long-wavelength and low-frequency limit.
Using linear response theory, the dielectric function $\varepsilon(\mathbf{q}, iw_n)$ can
be related to the density-density correlation function 
$
C(\mathbf{q}, \tau) = \langle T_\tau n(\mathbf{q}, \tau) n(-\mathbf{q}, 0)  \rangle
$
and to ${\widetilde{J}(\mathbf{q})}$
via
\begin{eqnarray}
\frac{1}{\varepsilon(\mathbf{q}, iw_n)} = 1 - \frac{\widetilde{J}(\mathbf{q})} {V_c} 
\int_0^{1/T} d\tau e^{i w_n \tau} C(\mathbf{q}, \tau),
\label{eqn:sumrule}
\end{eqnarray}
where $n(\mathbf{q}, \tau)$ is the density operator.
In the long-wavelength and low-frequency limit,
$C(\mathbf{q}, \tau)$ is directly related to $\kappa(T)$ via the
compressibility sum rule~\cite{pines}.
Therefore, a divergent dielectric function occurs when 
$
\varepsilon(\mathbf{q} \to 0, iw_n \to 0) = 0
$
leading to 
$
1 + \widetilde{J} (\mathbf{q} \to 0) (\partial N / \partial \mu)_T = 0,
$
which is identical to Eq.~(\ref{eqn:tc-first}).
This relation can be applied to both Bose and Fermi systems.
Next, we discuss $T_{FE}$ for a weakly interacting ($a \ll \lambda$) 
and dilute ($na^3 \ll 1 $) gas of Bose and Fermi molecules, 
where $a$ is the scattering and $\lambda$ is the thermal length.

{\it Fermi molecules:} As a first application of Eq.~(\ref{eqn:tc-general}), we analyze $T_{FE}$ 
for a weakly interacting gas of Fermi molecules at any $T$.
In this case, the molecular density is given by
$
n = 2F_{3/2}(z) [1 - 2 F_{1/2}(z)a/ \lambda_F]/\lambda_F^3
$
where $\lambda_F = [2\pi/(M T)]^{1/2}$ is the thermal length,
and leading to
$
\kappa(T) \approx [F_{1/2}(z)/F_{3/2}(z) - 2 F_{-1/2}(z) a/\lambda_F]/(n T)
$
where $0 \le z = \exp(\beta \mu) \le \infty$ is the fugacity, and
$
F_\nu(z) = [1/\Gamma(\nu)] \int_0^\infty x^{\nu - 1} dx / [z^{-1} e^x + 1]
$
is the Fermi integral.
Here, $\Gamma(\nu)$ is the Gamma function.
Thus, we obtain
\begin{equation}
T_{FE} = \frac{2|\mathbf{p}|^2} {3\varepsilon_0} 
\left[ \frac{F_{1/2}(z_c)} {F_{3/2}(z_c)} - \frac{2a}{\lambda_F} F_{-1/2}(z_c) \right]n,
\label{eqn:tc-fermi}
\end{equation}
where $z_c = \exp(\mu/T_{FE})$.
Notice that, in the classical ($z_c \ll 1$) limit, Eq.~(\ref{eqn:tc-fermi}) reduces to
$
T_{FE} = 2 |\mathbf{p}|^2 (1 - 2 z_c a/\lambda_F) n / (3 \varepsilon_0),
$
which shows that $T_{FE} \propto n$ for a gas of classical electric dipoles.
The $T_{FE}$ for an ideal (non-interacting) gas 
of Fermi molecules can be obtained by setting $a = 0$. 
For an ideal gas, when $T_{FE}$ is much smaller than the 
Fermi energy $\epsilon_F$, we obtain
$
T_{FE} \approx (2\sqrt{3} / \pi) \epsilon_F [|\mathbf{p}|^2 n/(\varepsilon_0 \epsilon_F) - 1]^{1/2},
$
which is valid for
$
|\mathbf{p}|^2 n/(\varepsilon_0 \epsilon_F) > 1
$
and
$
[|\mathbf{p}|^2 n/(\varepsilon_0 \epsilon_F) - 1]^{1/2} \ll 1.
$
For $T > T_{FE}$ a PFL phase exists and for $T < T_{FE}$ a 
FFL phase appears as shown in Fig.~\ref{fig:one}(a).

{\it Bose molecules for $T \ge T_{BEC}$:} 
As a second application of Eq.~(\ref{eqn:tc-general}), we analyze $T_{FE}$ 
for a weakly interacting gas of Bose molecules when $T \ge T_{BEC}$.
In this case, the molecular density is given by
$
n = B_{3/2}(z) [1 - 4 B_{1/2}(z) a / \lambda_B]/\lambda_B^3
$
where $\lambda_B = [1/(2\pi M T)]^{1/2}$ is the thermal length,
and leading to
$
\kappa(T) \approx [B_{1/2}(z)/B_{3/2}(z) - 4 B_{-1/2}(z) a / \lambda_B] / (n T)
$
where $0 \le z = \exp(\beta \mu) \le 1$ is the fugacity, and
$
B_\nu(z) = [1/\Gamma(\nu)] \int_0^\infty x^{\nu - 1} dx / [z^{-1} e^x - 1]
$
is the Bose integral.
Thus, we obtain
\begin{equation}
T_{FE} = \frac{2|\mathbf{p}|^2} {3\varepsilon_0} 
\left[ \frac{B_{1/2}(z_c)} {B_{3/2}(z_c)} - \frac{4a}{\lambda_B} B_{-1/2}(z_c) \right]n.
\label{eqn:tc-bose1}
\end{equation}
Notice that, in the classical ($z_c \ll 1$) limit, Eq.~(\ref{eqn:tc-bose1}) reduces to
$
T_{FE} = 2 |\mathbf{p}|^2 (1 - 4 z_c a/\lambda_B) n / (3 \varepsilon_0),
$
which again leads to $T_{FE} \propto n$ for a gas of classical electric dipoles.
The $T_{FE}$ for an ideal (non-interacting) gas 
of Bose molecules can be obtained by setting $a = 0$. 
Notice that, the second terms in Eqs.~(\ref{eqn:tc-fermi}) and~(\ref{eqn:tc-bose1}) 
are different by a factor of $2$ due to the degeneracy of 
pseudo-spin-$1/2$ fermions in contrast to pseudo-spin-$0$ bosons.

{\it Bose molecules for $T \le T_{BEC}$:} 
As a third application of Eq.~(\ref{eqn:tc-general}), we analyze $T_{FE}$ 
for an ideal (non-interacting) gas of Bose molecules when $T \le T_{BEC}$.
In this case, the molecular density is given by
$
n = B_{3/2}(z) / \lambda_B^3 + n_s
$
where $n_s = z / [V_c (1 - z)] = \alpha(T) n$ is the density of bosons 
in the condensed (zero-energy) state.
Here, $\alpha(T) = 1 - (T/T_{BEC})^{3/2}$ where
$
T_{BEC} = 2\pi [n / \zeta(3/2)]^{2/3} / M
$
is the critical BEC temperature for 
non-interacting dilute bosons, and $\zeta(x)$ is the Zeta function.
In this case, 
$
\kappa(T) = [B_{1/2}(z)/\lambda_B^3 + V_c n_s^2/z]/ (n^2 T),
$
which diverges in the thermodynamic limit~\cite{pathria} 
when $\{ N, V_c \} \to \infty$ but $n = N/V_c$ is a constant.
Thus, we obtain
\begin{eqnarray}
T_{FE} = \frac{2|\mathbf{p}|^2} {3\varepsilon_0} \left[ \frac{B_{1/2}(z_c)}{B_{3/2} (z_c)} (n - n_s) + 
\frac{V_c n_s^2}{z_c} \right],
\label{eqn:tc-bose2}
\end{eqnarray}
which is always smaller than $T_{BEC}$, and reduces to the non-interacting
limit of Eq.~(\ref{eqn:tc-bose1}) in the absence of BEC.

The two cases of Bose molecules allows the construction of the 
phase diagrams indicated in Fig.~\ref{fig:one}(b) and~\ref{fig:one}(c), respectively, 
where the PBL, FBL, PBEC and FBEC are identified
depending on the existence of a spontaneous average electric 
polarization $\langle {\bf P} \rangle$ and/or of a BEC fraction $\langle \phi \rangle$. 
The ferroelectric superfluid phases proposed here may be experimentally observed with 
currently available cooling techniques only when $T_{FE}$ is large enough. This requirement
imposes a condition on the size of the electric dipole moments of the molecules, and 
it is discussed next for the Fermi-Fermi Bose molecules.

{\it Fermi-Fermi Bose molecules:}
To set the scale, we consider the specific example of Li-K molecules consisting of
$^6$Li and $^{40}$K atoms in their ground state, where $|\mathbf{p}| = 3.6$ Debye~\cite{aymar}. 
We also choose an equal population mixture of $^6$Li and $^{40}$K atoms with parameters
$N = 10^5$ and $V = 10^{-7}$ $\rm{cm}^3$
leading to $T_{BEC} \approx 0.099 \epsilon_F$ and 
$T_{FE} \approx 21 T_{BEC}$. However, 
for a molecule with $|\mathbf{p}| = 1.0$ Debye, $T_{FE} = 1.6 T_{BEC}$.
Here, $\epsilon_F = k_F^2 / (2 m_r)$ is the Fermi energy, where 
$m_r$ is half of the reduced mass of Li and K atoms, 
and $k_F$ is the Fermi momentum with $n = k_F^3 / (6 \pi^2)$.
\begin{figure} [htb]
\centerline{\scalebox{0.5}{\includegraphics{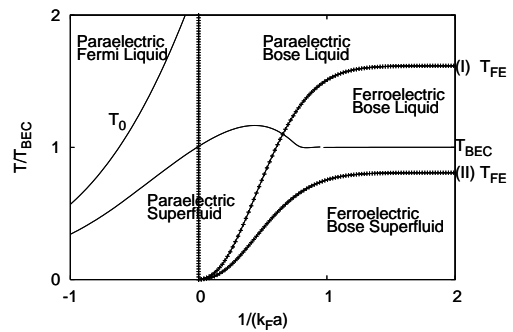} }}
\caption{\label{fig:two}
Schematic phase diagram of temperature $T/T_{BEC}$ versus scattering parameter $1/(k_F a)$
for an equal mixture of two-species Fermi-Fermi mixtures.
}
\end{figure}

For the two-species Fermi-Fermi mixtures (such as $^6$Li and $^{40}$K), 
we propose a schematic phase diagram of temperature $T$ versus 
scattering parameter $1/(k_F a)$ as shown in Fig.~\ref{fig:two}.
In the diagram, $T_0$ is the pair formation scale and Cooper pairs break above 
this temperature. In addition, a two-body bound state does not exist for negative $a$.
Thus, a polar bosonic molecule with dipole moment $|\mathbf{p}| \ne 0$ 
may form when both $T < T_0$ and $a > 0$ criteria are satisfied.
This also implies that a finite polarization $\mathbf{P}_0$ may exist below
$T < T_{FE}$ in the phase space where $T < T_0$ and $a > 0$ 
leading to a possible ferroelectric phase.
Depending on the relative values of $T_{FE}$ and $T_{BEC}$, we sketched two cases:
(I) $T_{FE} > T_{BEC}$ and (II) $T_{FE} < T_{BEC}$ which are shown as
dotted lines. Notice that as $1/(k_F a)$ decreases, the size of the molecules increases 
and $\mathbf{p}$ gets smaller, thus producing a reduced $T_{FE}$. 
Therefore, an equal mixture of $^6$Li and $^{40}$K correspond to case I
in Fig.~\ref{fig:two}. However, other mixtures with smaller dipole moments and/or densities
may correspond to case II. Next, we discuss trapped molecules.

{\it Heteronuclear molecules in a trap:} 
The trapping potential of polar molecules is given by 
$V ({\bf r}) = -\alpha |{\bf E} ({\bf r})|^2 - {\bf p} \cdot {\bf E} ({\bf r})$,
which are due to the coupling of the laser electric field ${\bf E} ({\bf r})$ 
with the induced and permanent dipole moments ${\bf p}_{ind} = \alpha {\bf E} ({\bf r})$ 
and ${\bf p}_{per} = {\bf p}$, respectively.
We assume that ${\bf E} ({\bf r}) = E_0 {\bf \hat x} \exp[-\rho^2/(2 w_\rho^2) -z^2/(2w_z^2)]$
has a gaussian profile controlled by the widths $w_\rho$ and $w_z$,
where $\rho^2 = x^2 + y^2$. 
Thus, the trapping potential can be approximated by   
$V ({\bf r}) = V_0 + M \Omega_{\rho}^2 \rho^2/2 + M\Omega_z^2 z^2/2$,
where $V_0 = - \alpha E_0^2 - |\mathbf{p}| E_0 \cos{\theta}$, and
$\Omega_{i} = ( 2 \alpha E_0 + |\mathbf{p}| E_0 \cos{\theta})/(M w_{i}^2)$
are the characteristic frequencies of the harmonic trap along $i = \rho, z$ directions.
Here, $\theta$ is the angle between ${\bf p}$ and ${\bf E} ({\bf r})$.
When $w_z \gg w_\rho$ ($w_z \ll w_\rho$) the trap is cigar (disc) shaped.

In the ground state of the system, the Thomas-Fermi approximation can be applied 
leading to
\begin{equation}
U n_0 ({\bf r}) = \tilde{\mu} - V ({\bf r})  + 
\int d{\bf r^\prime} J ({\bf r}, {\bf r^\prime}) n_0 ({\bf r^\prime}),
\end{equation}
and the electric polarization density is given by 
${\bf P} ({\bf r}) ={\bf  p} n_0 ({\bf r})$.
The integral equation can be solved analytically when 
$\beta = [|\mathbf{p}|^2/(4\pi \varepsilon_0)]/ U \ll 1$,
corresponding to a small ratio between the characteristic electric dipolar 
energy $|\mathbf{p}|^2/(4\pi \varepsilon_0 V_c)$ and the characteristic contact interaction energy $U/V_c$.
To zeroth order in $\beta$, the electric polarization is
${\bf P} ({\bf r}) = {\bf P}_{TF} ({\bf r}) = {\bf p} n_{TF} ({\bf r})$, where 
\begin{equation}
n_{TF} ({\bf r}) = n_{max} \left[ 1 - \rho^2/\rho_{c}^2 - z^2/z_c^2 \right].
\end{equation}
Here,  $n_{max} = ({\widetilde \mu} - V_0)/U$, 
$\rho_{c}^2 = ({\widetilde \mu} - V_0) / (M\Omega_{\rho}^2)$ and 
$z_{c}^2 = ({\widetilde \mu} - V_0) / (M\Omega_{z}^2)$.
The correction to first order in $\beta$ for the cigar shaped trap is
\begin{equation}
\delta {\bf P} ({\bf r}) = {\bf p} \beta \frac{4\pi}{18} {\bar n} \frac {\rho_{c}^2} {z_c^2}
\left[ c_1 - c_2 (\rho^2/\rho_{c}^2 - z^2/z_c^2) \right],
\end{equation}
where $c_1$ and $c_2$ are numerical coefficients, and ${\bar n} = N/V_c$ is the average molecular density.
Since ${\bf E} ({\bf r})$ is much stronger than the local electric fields produced by
electric dipolar interactions, the net polarization ${\bf P} ({\bf r}) = {\bf P}_{TF} ({\bf r}) + \delta {\bf P} ({\bf r}) $
points along ${\bf E} ({\bf r})$.

\begin{figure} [htb]
\centerline{\scalebox{0.3}{\includegraphics{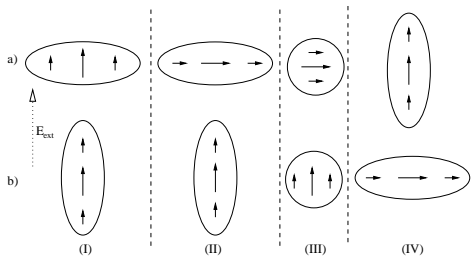} }}
\caption{\label{fig:three}
Direction of the polarization vector during time of flight for
a ferroelectric state,
when the external electric field is
a) perpendicular, and
b) parallel 
to the easy axis of polarization.
The clouds in each frame are not drawn to scale.}
\end{figure}

To distinguish between the fully polarized paraelectric superfluid and the true ferroelectric superfluid, 
we propose two experiments, as shown schematically in Fig.~{\ref{fig:three}},
where ${\bf E} ({\bf r})$ is turned off between times I and II.
In case (a), the external electric field and the easy axis for polarization are perpendicular. 
For the ferroelectric state, 
the polarization flips between times I and II (within nanoseconds to microseconds) 
towards the axial direction which corresponds to the easy axis for the 
electric polarization. This flip produces a short pulse of 
coherent dipolar radiation proportional to $N^2$. However, for the
fully polarized paraelectric state the polarization relaxes and reduces to zero within microseconds,
producing a longer pulse of incoherent dipolar radiation proportional to $N$.
A second coherent pulse may occur in time of flight
for the ferroelectric state as the anisotropy inversion in the cloud expansion takes place between
times III and IV, and causes a second flip of the electric polarization towards the new easy axis.
In case (b), the external electric field and the easy axis for polarization coincide. 
For the ferroelectric state, the polarization remains along the easy axis between times I and II, 
and between times III and IV when the anisotropy inversion occurs, 
the polarization flips towards the new easy axis with 
emission of coherent dipolar radiation (proportional to $N^2$). 
In this case only one pulse (within nanoseconds
to microseconds) should be observed after a few miliseconds of expansion.
For the paraelectric state similar results as in case (a) apply.

In conclusion, we analyzed the possibility of a ferroelectric transition in
Bose-Bose, Bose-Fermi or Fermi-Fermi heteronuclear molecules.
This transition is characterized by the appearance of a spontanous electric polarization 
below a critical temperature where the dielectric function diverges.
We obtained the order parameter equation, evaluated the
transition temperature and the electric polarization for ultracold heteronuclear 
(Bose or Fermi) molecules.
We discussed the existence of a ferroelectric Fermi liquid phase for polar Fermi molecules,
and the existence of a ferroelectric superfluid phase for polar Bose molecules
characterized by the coexistence of ferroelectric and superfluid orders.
We also proposed an experiment to detect ferroelectric correlations via the observation
of coherent dipole radiation pulses during time of flight. 
We thank NSF (DMR-0304380) for support.

\end{document}